# Wideband Optical Filters with Small Gap Coupled Subwavelength Metal Structures

Boyang Zhang, Junpeng Guo,* Robert Lindquist, and Stuart Yin

*Abstract*—In this letter, we show that the bandwidth of optical band-stop filters made of subwavelength metal structures can be significantly increased by the strong plasmonic near-field coupling through the corners of the periodic metal squares. The effect of small gap coupling on the spectral bandwidth is investigated by varying the gap size between the metal squares. An equivalent transmission line model is used to fit the transmission and reflection spectra of the metal filters. The transmission line model can characterize well the metal structures with the gap size larger than the near-field decay length. However, it fails to model the transmission and reflection spectra when the gap size reaches the decay range of the near-field in the small gaps.

## I. Introduction

USING subwavelength metal structures as infrared optical filters can be dated back to the 1960s [1]-[7]. Periodic metal plates function as band-stop filters in the infra-red. Metal-mesh structures exhibit the opposite property and function as band-pass infra-red optical filters. Recently, there is an increasing interest on localized surface plasmon coupling between metal nano-particles [8]-[10]. It has been found that the near-field plasmonic couplings can change the resonance spectra of the coupled systems. One significant feature of the small gap coupled metal structures is the broadening of the resonance spectra. Previous works have mainly focused on the coupling in metal nano-particle pairs and its influence on the spectra of the scattered field. In this paper, we propose a coupled metal structure in which the metal elements are coupled through the small gaps between the corners. The motivation of this research is to investigate the feasibility of making wideband infra-red optical filters using the small gap coupled metal structures. The coupled structure is different from the other previously reported structures in that all adjacent metal elements are coupled through near-fields equally.

## II. Coupled Metal Structure and Equivalent Transmission Line Model

Manuscript received on September 2, 2011. The work was partially supported by the National Science Foundation (NSF) under the award NSF-0814103 and the National Aeronautics and Space Administration (NASA) under the grant NNX07 AL52A.

Boyang Zhang, Junpeng Guo, and Robert Lindquist are with the Department of Electrical and Computer Engineering, University of Alabama in Huntsville, Huntsville, AL 35899. Stuart Yin is with the Department of Electrical and Computer Engineering, Pennsylvania State University, University Park, PA 16802. *Correspondence should be addressed to Dr. Junpeng Guo via email : guoj@uah.edu..

Fig. 1 (a) shows the top view of the coupled metal structure optical filter. Two gold metal squares with thickness 20 nm are placed in each unit cell on a glass substrate. The length of the squares' sides is $a$ and the period is $\Lambda$. The gap between the adjacent metal squares in the horizontal and vertical directions is $g = (\Lambda-2a)/2$. The structure is polarization independent at normal incidence.

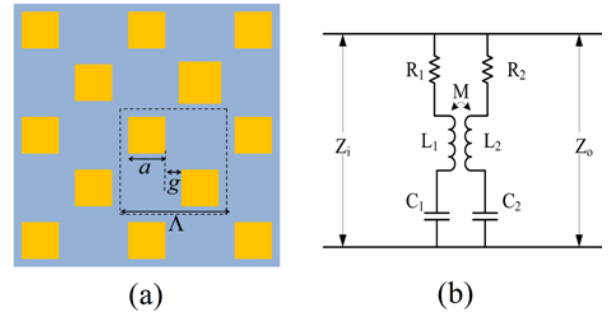

Fig. 1. (a) Schematic drawing of the coupled metal structure. (b) The equivalent transmission line model of the coupled metal structure.

The equivalent transmission line model has been widely used to model metal structure infra-red optical filters [1]-[4], [6]. When the excitation frequency reaches the order of THz, electrons in metals start to cause the inductance in metals. Regular metal plate array can be described by a series RLC resonator with L representing the inductance, C representing the capacitance, and R representing the loss [11]. When two inductors are close to each other, the coupling between them can be described by mutual inductance M. Based on this principle, an equivalent transmission line model for the coupled metal structure is proposed and shown in Fig. 1 (b). Depending on the gap size between the metal plates, two types of plasmonic couplings can occur: near-field coupling and far-field coupling [12]. Near-field coupling occurs when the metal elements are brought within the decay range of the near-field of the plasmon modes. Far-field coupling is caused by the radiation interactions between the periodic metal squares.

## III. Results and Discussions

### A. Regular metal arrays

First, the transmission and reflection spectra of regular periodic metal structures with only one metal square in the unit cells are calculated. In the simulations, the metal squares have the size $a$ = 600 nm and the period is varied from 1200 nm to 1500 nm. The calculations were carried out by using the finite element method (Comsol Multiphysics). The Drude model for the electric permittivity of the gold is used in our calculations. The solid lines in Fig. 2 show the calculated spectra. Both the reflection and transmission show strong resonances due to the surface plasmon resonances in the metal arrays [13]. At the resonance, the transmission reaches zero. As the period increases, the resonance wavelength is red-shifted indicating that the surface plasmon resonance wavelength in the metal arrays increases as the period increases. This trend is similar to the spectral red-shift that occurs in the periodic metal hole arrays when the period of holes increases [13].

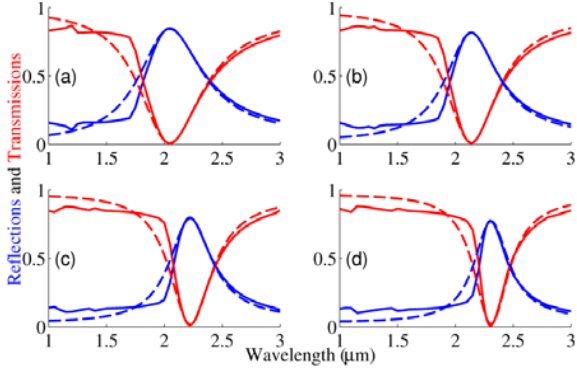

Fig. 2. The transmission and reflection spectra of regular metal square arrays: (a) Λ = 1200 nm, (b) Λ = 1300 nm, (c) Λ = 1400 nm, (d) Λ = 1500 nm. The solid curves are from numerical simulations and the dashed curves are from the equivalent transmission line circuit model.

The transmission line model is then used to fit the numerically calculated transmission and reflection spectra. The transmission line model is a series RLC circuit, in which we set $R_2$ is set to $\infty$ in Fig. 1 (b). The transmission line model results are shown in Fig. 2 as the dashed line curves. The resonance of the transmission line mode depends on the RLC resonator with the product LC determining the resonance frequency, L/C determining the bandwidth, and the resistance R determining the absorption. The transmission line model is unique and each structure can only be modeled by one set of R, L, C parameters. The best fitted values of $R_1$, $L_1$, and $C_1$ are 17 Ω, 5.710×10⁻¹³ H, and 2.066×10⁻¹⁸ F for Λ = 1200 nm; 21 Ω, 7.709×10⁻¹³, and 1.667×10⁻¹⁸ F for Λ = 1300 nm; 24 Ω, 1.052×10⁻¹² H and 1.324×10⁻¹⁸ F for Λ = 1400 nm; and 27 Ω, 1.406×10⁻¹² H, 1.067×10⁻¹⁸ F for Λ = 1500 nm. The spectra of the transmission line circuit model show the exact *Lorentzian* shape. The spectra of the numerical simulations deviate from the transmission line model in the short wavelength region because the optical property of gold is highly dispersive in the short wavelength regime and the RLC parameters are independent of the wavelength.

### B. Coupled structures

The simulated spectra of coupled metal structures where two metal elements are included in each unit cell are shown in Fig. 3. Again, the metal squares have the size $a$ = 600 nm. The period Λ varies from 1200 nm to 1500 nm and the gap size changes from 0 to 150 nm. In Fig. 3 (b)-(f), the small gap coupled structures have significantly increased bandwidth compared to those of regular metallic arrays with same square size and period shown in Fig. 2. As the *gap* becomes smaller, the resonance bandwidth becomes broader. Also, as the gap is reduced from 150 nm to 50 nm (Fig. 3 (f) to (d)), the resonance shows slight blue-shift. However, a further decrease from 50 nm to 10 nm (Fig. 3 (d) to (b)) leads to a slight red-shift. In this case, the strong near-field coupling plays the significant role in increasing the resonance bandwidth as shown in Fig. 3 (a) shows the spectra when the metal squares are in contact at the corners (Λ = 2a =1200 nm). We see no obvious bandwidth widening compared to Fig. 2 (a) but a large shift to the shorter wavelength and narrower bandwidth compared to Fig 3 (b)-(f). This disruptive change compared to spectral variation from Fig. 3(b) to (f) is due to the new plasmon mode when metal elements are in contact and is different from plasmonic couplings [14].

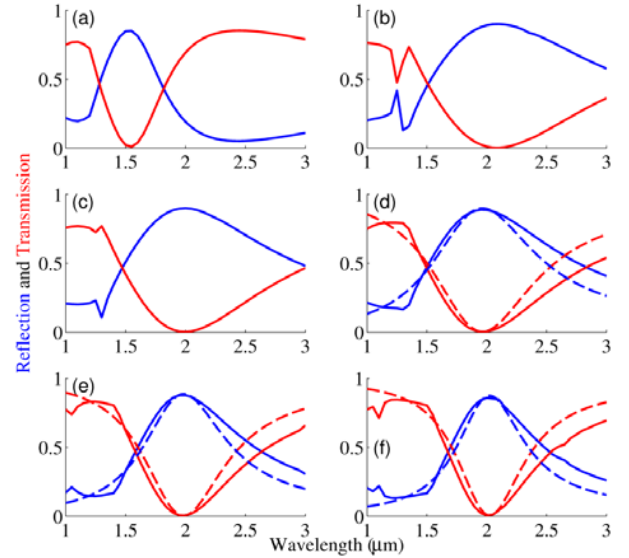

Fig. 3. The transmission and reflection spectra of coupled metal structures. (a) Λ = 1200 nm ($g$ = 0 nm), (b) Λ = 1220 nm ($g$ = 10 nm), (c) Λ = 1250 nm ($g$ = 25 nm), (d) Λ = 1300 nm ($g$ = 50 nm), (e) Λ = 1400 nm ($g$ = 100 nm), (f) Λ = 1500 nm ($g$ = 150 nm). The solid curves are from numerical simulations. In (d)-(f), the dashed curves are from the transmission line model.

In an investigation into the relation between resonance wavelength and the gap size, the simulation indicates that the resonance wavelength red-shifts as the gap ($g$) decreases from 40 nm to 10 nm where the strong near-field coupling is dominant. When the gap size decreases from 150 nm to 50 nm, the resonance wavelength shows a decreasing trend due to the far-field couplings. We measure the decay range of enhance near-field on simulated electric field distributions (shown in Fig. 4) and find it is about 50 nm.

The dash lines in Fig. 3(d)-(f) are the transmission line model results. When the gap size is large, good agreement is found between the transmission line model and the numerical simulations. The RLC parameters are the same as used in Fig. 2 for each period and the fittings are obtained by tuning the coupling factor $M$. Only one value of $M$ can be found to achieve the good fitting for each gap size. The best fitted $M$ values are -1.15×10⁻¹³ H in Fig. 3(d), -2.15×10⁻¹³ H in Fig. 3(e), and -3.25×10⁻¹³ H in Fig. 3(f). For structures with smaller gaps, the resonance spectra have even broader spectral width as shown in Fig. 3 (b) and (c). However, due to the red-shift of the resonance, we are not able to find appropriate value of $M$ to achieve transmission line model fitting for structures with small gaps. When the corners between adjacent metal squares are close, additional capacitance appears in the gaps. The increased capacitance as the gap decreases causes the red-shift of the resonance spectra. Therefore, spectral fitting cannot be achieved by simply tuning $M$ when the gap is small. This phenomenon can also be explained by the coupled dipole theory [15]. For structures with larger gaps, near-field couplings are eliminated. The far-field coupling is linearly related to the gap size. When the gap is smaller than the decay range of the near-field, strong nonlinear near-field coupling through corners starts to play the dominant role. In the transmission line model, the coupling between RLC resonators is linear coupling of two harmonic oscillators. Therefore, the transmission line model can fit the far-field coupling well, but cannot mimic the near-field coupling.

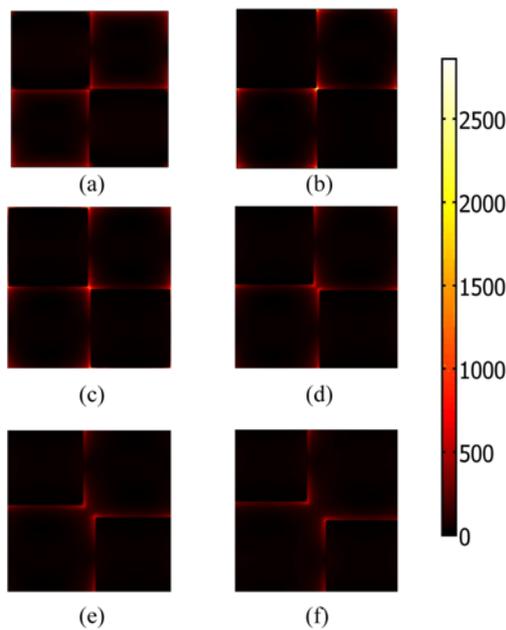

Fig. 4. Simulated electric field intensity $|E|^2$ distributions of coupled metal structures at their resonance wavelengths for different gap sizes: (a) $g = 0$ nm, (b) $g = 10$ nm, (c) $g = 25$ nm, (d) $g = 50$ nm, (e) $g = 100$ nm, (f) $g = 150$ nm.

In order to understand the couplings in the structures, we plot simulated electric field intensity $|E|^2$ distributions for different gap sizes in Fig. 4. For small gaps, strong enhanced optical fields can be seen inside the gaps as shown in Fig. 4 (b) and (c). The enhanced electric fields in the gaps indicate strong near-field optical couplings between the two metal squares. This near-field coupling results in wider spectral band than that in the large gap structure. When the gap size is larger than 50 nm, the near-field interaction becomes weaker and the resonance spectral band of the metal structure becomes less broad.

## IV. Conclusion

In summary, we have calculated the transmission and reflection spectra of small gap coupled metal square array structures. It was found that when the gap between the metal squares becomes smaller, the resonance bandwidth becomes wider. The increased bandwidth is due to the near-field coupling via the corners of the metal squares. An equivalent transmission line model was developed to describe the coupled metal structure. The spectra of the metal structure with the gap size larger than the near-field decay range can be characterized accurately using the transmission line model. When the gap becomes small, the near-field coupling dominates and the transmission line circuit model cannot fit the spectra of the metal structure filter.